# KiNet: A Deep Neural Network Representation of Chemical Kinetics

Weiqi Ji, Sili Deng*

Department of Mechanical Engineering, Massachusetts Institute of Technology,
Cambridge, MA 02139

*Corresponding Author: silideng@mit.edu (Sili Deng)

**Abstract**

Deep learning is a potential approach to automatically develop kinetic models from experimental data. We propose a deep neural network model of KiNet to represent chemical kinetics. KiNet takes the current composition states and predicts the evolution of the states after a fixed time step. The long-period evolution of the states and their gradients to model parameters can be efficiently obtained by recursively applying the KiNet model multiple times. To address the challenges of the high-dimensional composition space and error accumulation in long-period prediction, the architecture of KiNet incorporates the residual network model (ResNet), and the training employs backpropagation through time (BPTT) approach to minimize multi-step prediction error. In addition, an approach for efficiently computing the gradient of the ignition delay time (IDT) to KiNet model parameters is proposed to train the KiNet against the rich database of IDT from literature, which could address the scarcity of time-resolved species measurements. The KiNet is first trained and compared with the simulated species profiles during the auto-ignition of $H_2$/air mixtures. The obtained KiNet model can accurately predict the auto-ignition processes for various initial conditions that cover a wide range of pressures, temperatures, and equivalence ratios. Then, we show that the gradient of IDT to KiNet model parameters is parallel to the gradient of the temperature at the ignition point. This correlation enables efficient computation of the gradient of IDT via backpropagation and is demonstrated as a feasible approach for fine-tuning the KiNet against IDT. These demonstrations shall open up the possibility of building data-driven kinetic models autonomously. Finally, the trained KiNet could be potentially applied to kinetic model reduction and chemistry acceleration in turbulent combustion simulations.

**Keywords**: Deep Learning; Chemical Kinetics; ResNet; BPTT; Ignition Delay Time

## 1. Introduction

Detailed chemical kinetic models are usually in the form of providing elementary reaction pathways among species. The determination of reaction pathways and the associated rate constants is time-consuming and often requires expert knowledge. The recent revolution of deep learning [1] has attracted a growing interest in automatically building kinetic models from experimental data with deep neural network (DNN) [2,3]. As initial attempts [2,3], Ranade et al. proposed to utilize the temporal evolution profiles of chemical species during the pyrolysis and oxidation processes in homogenous reactors as the targets. In computations, the evolution of species profiles can be described by a system of ordinary differential equations (ODEs)

$$\frac{d\boldsymbol{\phi}}{dt} = S(\boldsymbol{\phi}), \tag{1}$$

where $\boldsymbol{\phi}$ is the composition state vector, including the species concentrations, temperature, and density. The reaction source terms $S(\boldsymbol{\phi})$ are determined by the adopted chemical kinetic model. The integrated equations over a time step $\delta t$ can be discretized as

$$\boldsymbol{\phi}(t + \delta t) = \boldsymbol{\phi}(t) + \boldsymbol{\delta\phi}(t), \tag{2}$$

where $\boldsymbol{\delta\phi}(t)$ is the residual and represents the overall change in composition states during the time step.

DNN models have been utilized to predict the evolution of the species profiles by predicting either $S(\boldsymbol{\phi})$ in the differential form as in Eq. (1) [2–4] or $\boldsymbol{\delta\phi}$ in the residual form as in Eq. (2) [5–11]. In the differential approach, the evolution of composition states is computed using ODE solvers, while the residual approach only involves the forward pass of DNN models via algebraic operations. Therefore, previous studies have explored the residual approach for tabulation in turbulent combustion simulations [5–11] to alleviate computational time and memory requirement. Despite the successful demonstrations of DNN models for chemical kinetics, the training of DNN models to capture the complexity and strong nonlinearity of chemical kinetics under a wide range of thermodynamic conditions remains challenging. Since a DNN model usually involves thousands and even millions of parameters, the training of DNN heavily relies on the efficient evaluation of the gradients of network outputs to network parameters using backpropagation.

In the differential approach that predicts the reaction source terms, the DNN model can be efficiently trained if the time-resolved measurements of all species concentrations, temperature and density are available. The time derivatives of these measurements can be readily utilized as the labels (targets) for the reaction source terms. However, obtaining time-resolved measurements for all composition states are generally intractable due to the limited availability of diagnostic techniques [12]. The DNN model can also be trained with partially observed species profiles, in which case only some of the species profiles are available and/or time-solved. The unknown/unresolved species profiles can be estimated via integrating Eq. (1) with an ODE solver from the known initial conditions, and the DNN model can be trained by minimizing the loss of the known/resolved species profiles. However, this approach would require the backpropagation of the species profiles across the ODE solver, which is still challenging for the large number of model parameters and stiff ODEs [13,14].

Conversely, in the residual approach that predicts the overall change of the composition states over time, efficient algorithms for backpropagation through time (BPTT) is readily available in popular deep learning frameworks, such as TensorFlow [15] and PyTorch [16]. Although previous explorations of the residual approach [5–11] focused on training DNN models for efficient tabulation in turbulent combustion simulations, the procedure can be directly applied to train models against experimental measurements of species profiles from homogenous reactors.

However, there are several challenges in applying the residual formula to build a data-driven kinetic model as elaborated below.

**(I)** Training a compact DNN model valid for a wide range of thermodynamic states is challenging due to the high-dimensionality, nonlinearity, and stiffness of chemical kinetic systems. To limit the size of DNN models and the size of training datasets, clustering techniques such as Self-Organizing Map (SOM) [17] have been employed to cluster the composition space into subdomains. Instead of training a humongous DNN to cover all thermodynamic states, several smaller DNNs are trained such that each DNN model represents the kinetics within each subdomain [11,18]. However, the clustering requires prior knowledge of the statistical distribution of the compositions, which is often unavailable from experimental data.

**(II)** DNN models trained via minimizing single-step prediction errors do not guarantee small errors in multi-step predictions. By recursively feeding the outputs from the last time step to the DNN as new inputs, the DNN could make long-period predictions. Blasco et al. [7] have shown that the multi-step prediction error could be significant while the single-step prediction error is very small. Consequently, the accumulated error over time may prevent the DNN from accurately reproducing the evolution of species profiles measured during pyrolysis and ignition processes. Moreover, reducing the error accumulation is also necessary for the subsequent application of the data-driven kinetic models in multi-dimensional combustion simulations.

**(III)** Time-resolved species profile measurements are very limited for DNN training. On the other hand, the ignition delay time (IDT) has been extensively reported in the literature since IDT can be readily determined from pressure traces. Furthermore, recent work has demonstrated automated measurements of IDT at high-throughput in a miniature shock tube, where each measurement only takes 4 seconds [19]. Therefore, training DNN from IDT would be desirable but challenging, which requires the efficient computation of the gradient of IDT with respect to the parameters of DNN.

In this work, we propose a kinetic neural network framework, termed as KiNet, to address the above challenges. Specifically, our work has the following novelties: (i) Residual Network (ResNet) [20] is adopted to increase the generalization performance of the DNN model. (ii) The multi-step prediction error is reduced by minimizing multi-step loss with BPTT approach [21], which is commonly used in training the Recurrent Neural Network (RNN). (iii) An approach is proposed to efficiently compute the gradient of IDT to the DNN model parameters based on the correlations between the gradient of IDT and that of temperature, which can be efficiently computed via backpropagation. Consequently, the KiNet shall open up the possibility of building a DNN representation of chemical kinetics without referring to the classical form of elementary reactions and Arrhenius formula.

This manuscript is structured as follows: Section 2 describes the architecture and training procedure of the proposed KiNet models. Section 3 demonstrates the performance of KiNet with simulated species profiles using a detailed kinetic model of hydrogen. Section 4 introduces an efficient approach to compute the gradient of IDT. Finally, the Conclusions summarize the key features of KiNet and provide an outlook for potential applications of KiNet.

## 2. The Architecture and Training of KiNet

2.1 Architecture of KiNet

Figures 1 a and b show schematic diagrams of the neural network models for the single-step and multi-step forward prediction of Eq. (2), respectively. For single-step prediction, the network takes the composition states ($\boldsymbol{\phi}(t_0)$: mole fractions of all species, temperature, and density) at the current time $t_0$, and outputs the composition states at the next time step $t_0 + \delta t$. More specifically, the residual $\boldsymbol{\delta\phi}(t)$ is predicted by the hidden layers of the ResNet, and the new composition states are obtained by adding an identity mapping of the initial states $\boldsymbol{\phi}(t_0)$ to the residual. Within the ResNet, fully connected layers are employed, where all of the activation functions are ReLU, and there is no activation function for the output layer.

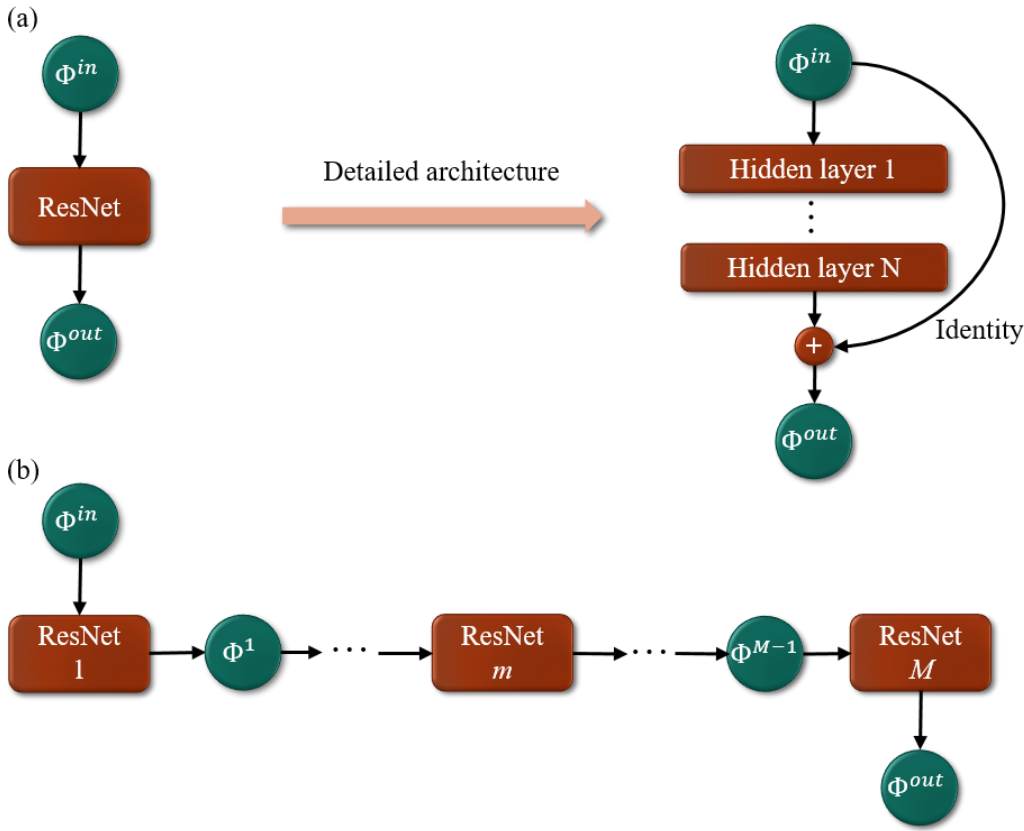

Figure 1. Schematics of the ResNet for (a) single-step prediction and (b) multi-step prediction. $\Phi^{in}$ and $\Phi^{out}$ correspond to the composition states at the current time step and next time step, respectively. All ResNets share the same weights.

To achieve multi-step prediction over $M$ steps, $M$ ResNets are stacked to form the KiNet, such that the $m$-th ResNet takes the states at $t_0 + (m-1)\delta t$ as inputs and outputs the states at $t_0 + m\delta t$. Although the single ResNet might contains only several hidden layers, say less than five, the KiNet is essentially a very deep neural network since the stacked network for *M*-step prediction contains $N*M$ hidden layers besides the output layers between any two adjacent ResNet blocks. Note that all ResNets are identical, i.e., they share the same weights, in order to make the model trainable.

The identity mapping in the ResNet architecture offers the following advantages: (i) The residuals are centered at zero which facilitates the training. (ii) The identity mapping could reduce the level of gradient explosion and vanishing problems in training very deep neural networks [20] such as KiNet. (iii) The ResNet can be viewed as a one-step method for the temporal integration of Eq. (1) [22–24], which makes the neural network model intuitively similar to the conventional chemical kinetic models. The similarity could potentially facilitate the knowledge transfer from the conventional chemical kinetic model to KiNet.

The activation function of ReLU has the following advantages compared to the hyperbolic tangent function (Tanh) used in previous studies [8,9,11]. (i) ReLU is less computationally expensive than Tanh since it involves simpler mathematical operations. (ii) The gradient of ReLU keeps constant for large inputs while the gradient of Tanh vanishes, so the adoption of ReLU alleviates the gradient vanishing problems for KiNet.

2.2 Training of KiNet

KiNet can be trained by minimizing the single-step prediction error. However, as shown by Blasco et al. [7], such approach cannot guarantee the performance of multi-step prediction. The deviation can be attributed to the compounding error during the recursive predictions that changes the distribution of the input for future prediction steps, breaking the common i.i.d (independent and identically distributed) assumption of train-test in supervised learning [25]. In addition, the gradient direction of the single-step prediction to model parameters is not necessarily the same as that of the multi-step prediction, such that minimizing the single-step prediction loss by gradient descent does not guarantee the minimization of the multi-step prediction loss.

Consequently, we design the loss of KiNet as the summation of the loss at each time step, i.e.,

$$\text{Loss} = \sum_{m=1}^{M} loss(\phi^{m,predict}, \phi^{m,label}). \tag{3}$$

In the current work, the gradient of the loss is efficiently computed via BPTT [21], and the BPTT is implemented in PyTorch with the dynamic computational graph [16]. In such a way, the multi-step prediction error can be greatly reduced. In the following discussion, the KiNet trained by minimizing M-step ($M \geq 1$) prediction loss is denoted as KiNet-M. Directly training KiNet over long time sequences is very time-consuming so we use the following incremental approach [23]. Initially, the network is trained by minimizing one-step prediction loss. Then, the trained model ResNet-1 serves as initialization for training ResNet-2, and the process goes on to eventually train KiNet-M. The incremental approach also prevents the gradient explosion induced by the large prediction error in the intermediate steps. The largest value for $M$ mainly depends on the affordability of the computational cost of the training. Since all the predictions of the intermediate hidden layers in the multi-step prediction are recorded in BPTT approach, both the computational time and memory requirements increase with $M$.

## 3. Demonstration of Predicting Auto-ignition Process

Next, we shall demonstrate the KiNet trained with simulated species profiles during the auto-ignition processes of the $H_2$/air mixtures. The ignition process is chosen for its central importance in developing chemical kinetic models, extensive experimental measurements in the literature, and practical relevance. The time-resolved profiles are generated from the simulation of constant pressure reactors using Cantera [26]. The initial mixture is $H_2$/air, with $O_2:N_2 = 1:3.76$ by volume. The initial thermodynamic conditions are randomly sampled from a wide range of pressures (1-20 atm), temperatures (900-1600 K), and equivalence ratios (0.5-2). The species profiles are collected up to the instance when the temperature is within 0.1 K of the equilibrium temperature or reaches 1 ms, whichever comes first. The cutoff time of 1 ms reduces the size of datasets for very long IDTs at low temperatures and speeds up the demonstration. IDT is defined as the instance when the temperature increases by 400 K compared to the initial temperature, and the recorded IDT ranges from 0.5-1000 µs. The adoption of open-source software Cantera and small dataset size make the demonstration easy to reproduce and could potentially serve as a kinetic MNIST (Modified National Institute of Standards and Technology) database [27] for future kinetic neural network studies. 4000 samples of initial conditions are drawn and split into training, validation and test datasets with the ratios of 70%, 15%, and 15%, respectively. The detailed reaction model from Li et al. [28] is adopted, which contains 9 species for the ignition of $H_2$/air mixtures: $H_2$, $O_2$, $H_2O$, $OH$, $H$, $O$, $HO_2$, $H_2O_2$, and $N_2$. Therefore, there are 11 inputs for KiNet, including the mole fractions of the nine species, temperature, and density.

Figure 2 shows a sample temporal evolution of species profiles from the training datasets. The time step is chosen to be $10^{-7}$ s. The principal of specifying the time step is to be as large as possible but small enough to maintain sufficient resolution near the ignition point. Each pair of composition states at the current time $t$ and the next time step $t + \delta t$ forms a training sample, and each ignition trajectory contains a large number of training samples. A resampling strategy is applied to balance the training samples from different initial conditions and reduce the size of the training dataset. For example, a maximum number of 1024 training samples are randomly selected from each ignition trajectory. Consequently, the database constructed in the current work involves ~$10^6$ samples for demonstrating the training and testing of the KiNet.

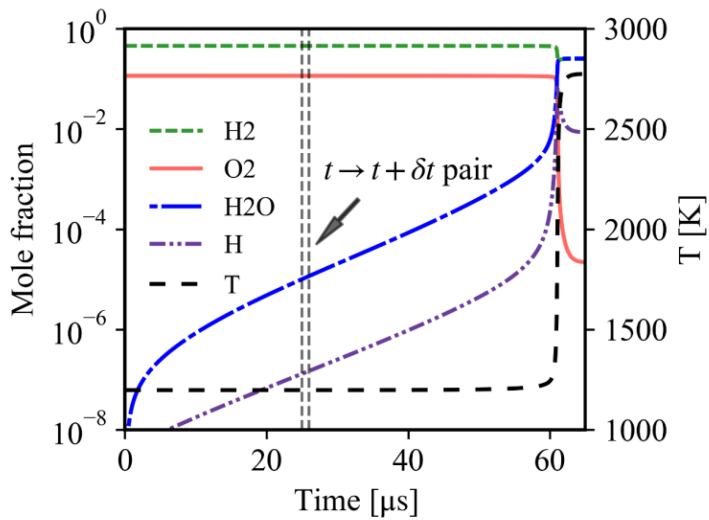

Figure 2. Sample temporal evolution of the species profiles during the auto-ignition process. The initial condition is 18.659 atm, 1198.426 K and equivalence ratio of 1.981. Each $t$ and $t + \delta t$ data pair in the time series is stored as a training sample.

The composition states are properly scaled to facilitate the training. As shown in Fig. 2, the species mole fractions of radicals are very low before ignition. Therefore, all of the inputs and outputs are transformed to log-scale. Then, they are centered and normalized by their mean and standard deviation via

$$\widetilde{\boldsymbol{\phi}} = \frac{\log(\boldsymbol{\phi}) - mean(\log(\boldsymbol{\phi}))}{std(\log(\boldsymbol{\phi}))}. \tag{4}$$

Mean square error (MSE) of the scaled outputs is employed as the loss function for each time step, and the loss function for M-step prediction is

$$\text{Loss} = \sum_{m=1}^{M} MSE\left(\widetilde{\boldsymbol{\phi}^{m,predict}}, \widetilde{\boldsymbol{\phi}^{m,label}}\right). \tag{5}$$

Each ResNet block contains three hidden layers with 100 neurons in each layer. The optimization algorithm is ADAM [29]. The learning rate is set to be 1e-3 and gradually decreased when training long sequences. The batch-size is set to be 128 and gradually increased as the portion of samples with large prediction error gradually decreases during the optimization. The final KiNet is trained up to 50-step prediction (5 µs) since it already well predicts species profiles and IDTs as shown below.

Figures 3 and 4 show the predicted temperature and mole fraction of H radical versus the label data of the simulated results with the $H_2$ model. Two representative trained models of KiNet-1 and KiNet-50 are presented. 20,000 samples are presented, and all the samples are drawn from the test datasets. KiNet-1 gives very good one-step forward predictions, as shown in Fig. 3a. However, it shows significant errors for 50-step forward predictions in Fig. 3b. The large errors are concentrated in the temperature range of 1500-2700 K, which correspond to the ignition period where the temperature increases sharply. Similar to the concurrence of the large prediction error and rapid temperature change, the large error in the predicted mole fraction of H radical is also in concordance with ignition as shown in Fig. 4b. Figures 3c-d and 4c-d show the predictions using KiNet-50. By minimizing multi-step prediction error, KiNet-50 significantly improves the 50-step forward predictions while maintains very good single-step forward predictions. Especially, some extremely ill-predicted scenarios with KiNet-1 shown in Figs. 3b and 4b are eliminated with KiNet-50.

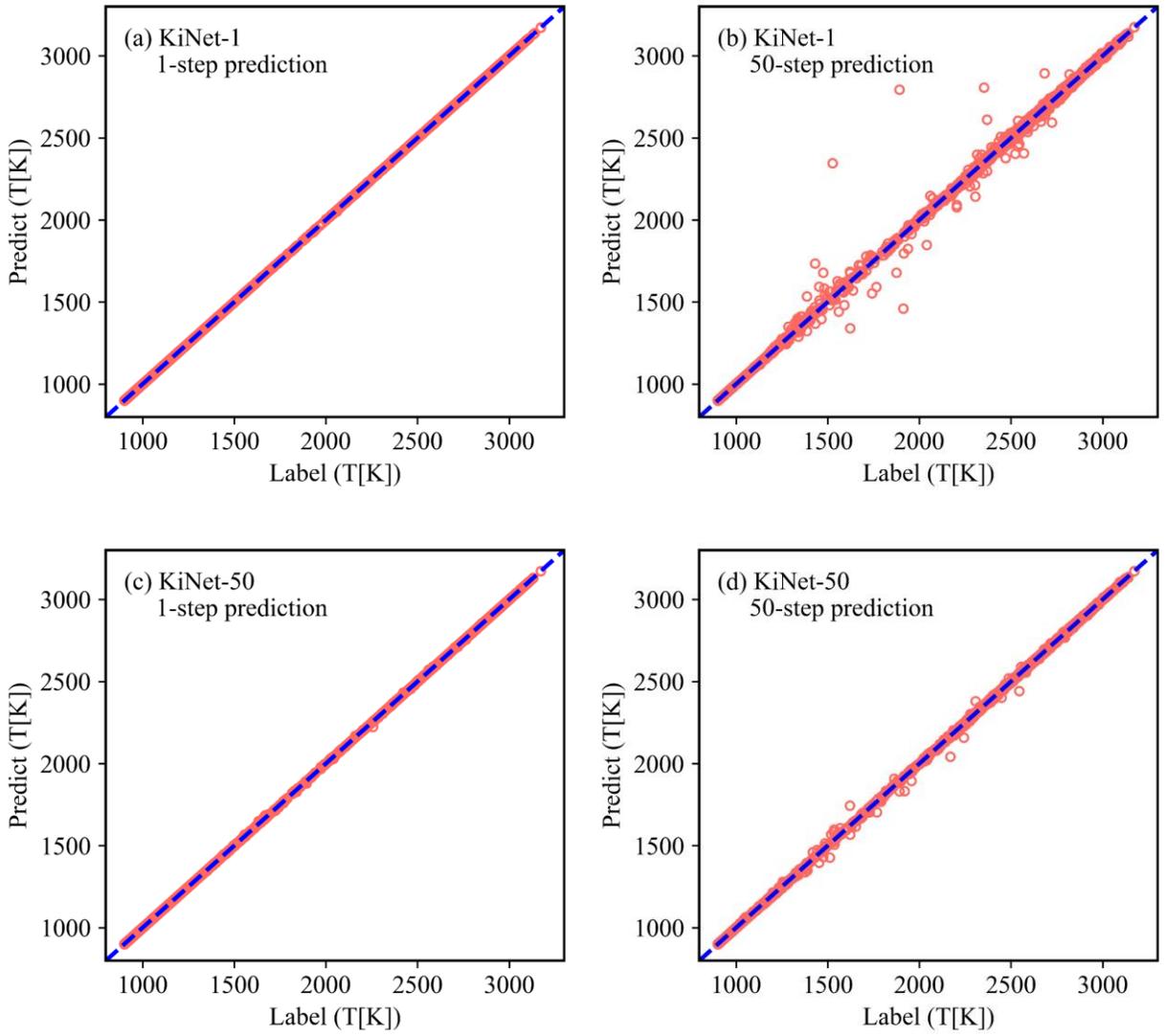

Figure 3. The predicted temperatures after one step and 50 steps versus corresponding label data using KiNet-1 and KiNet-50. 20,000 samples are presented.

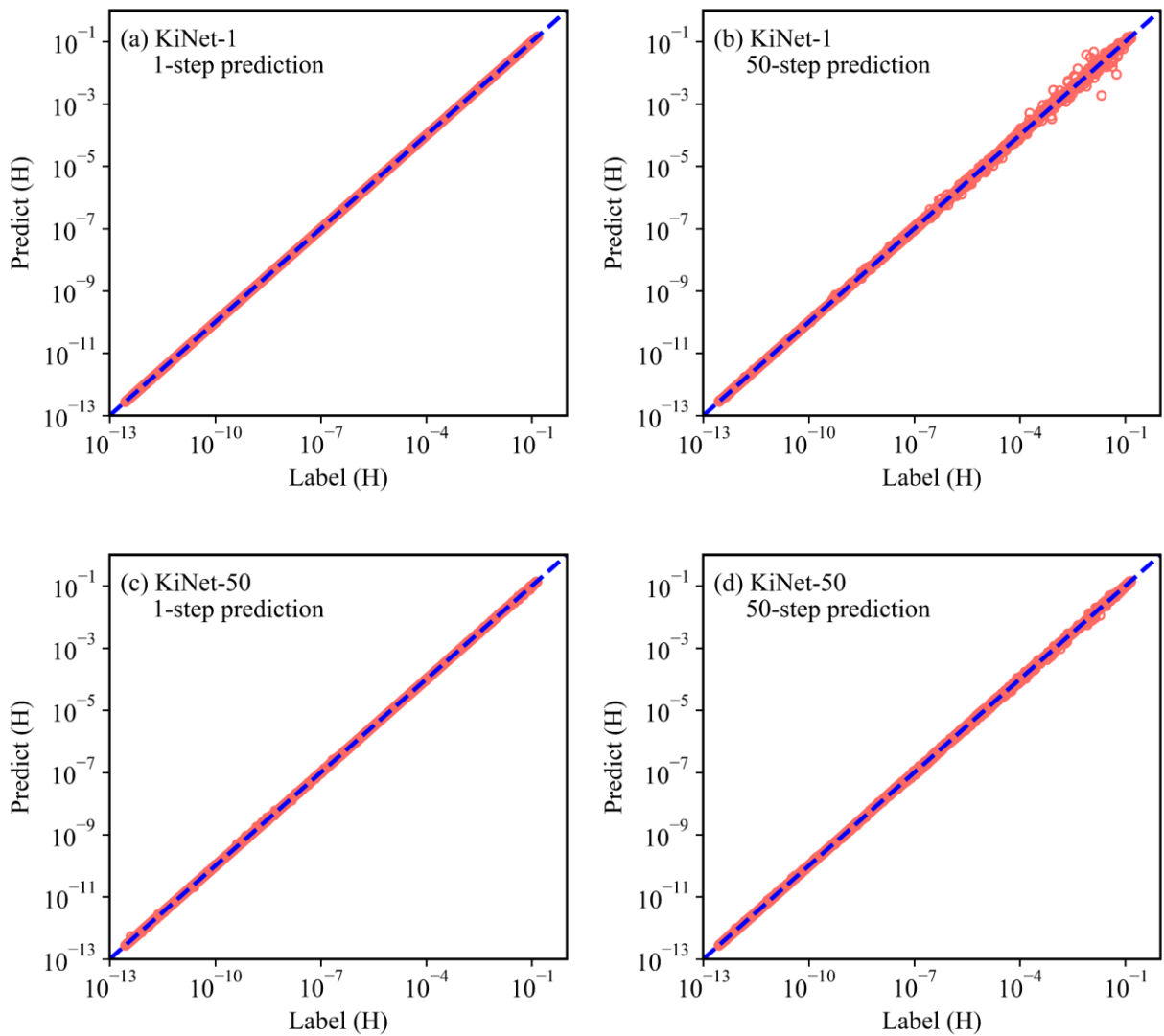

Figure 4. The predicted mole fractions of H radical after one step and 50 steps versus corresponding label data using KiNet-1 and KiNet-50. 20,000 samples are presented.

The KiNet-50 is then recursively applied to predict the species profiles over the entire ignition process. A sample evolution of the species mole fractions and temperature predicted using KiNet-50 is presented in Fig. 5. It can be seen that the KiNet-50 can accurately predict the temperature and species profiles, and hence, the IDT. Moreover, the predicted IDTs for the entire test datasets of 600 initial conditions using KiNet-1 and KiNet-50 are shown in Fig. 6. In general, both KiNet-1 and KiNet-50 give reasonably good predictions of IDT; however, KiNet-1 shows large error when IDT is longer than 1e-4 s. On the contrary, KiNet-50 significantly reduces the error for those cases with long IDT.

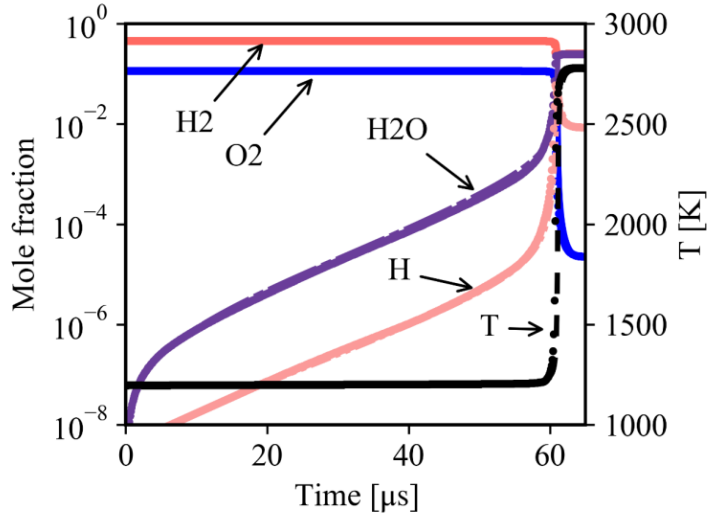

Figure 5. Sample time evolutions of temperature and species profiles simulated with the chemical kinetic model (label, shown as lines) and predicted with KiNet-50 (prediction, shown as symbols). The initial conditions are the same as in Fig. 2.

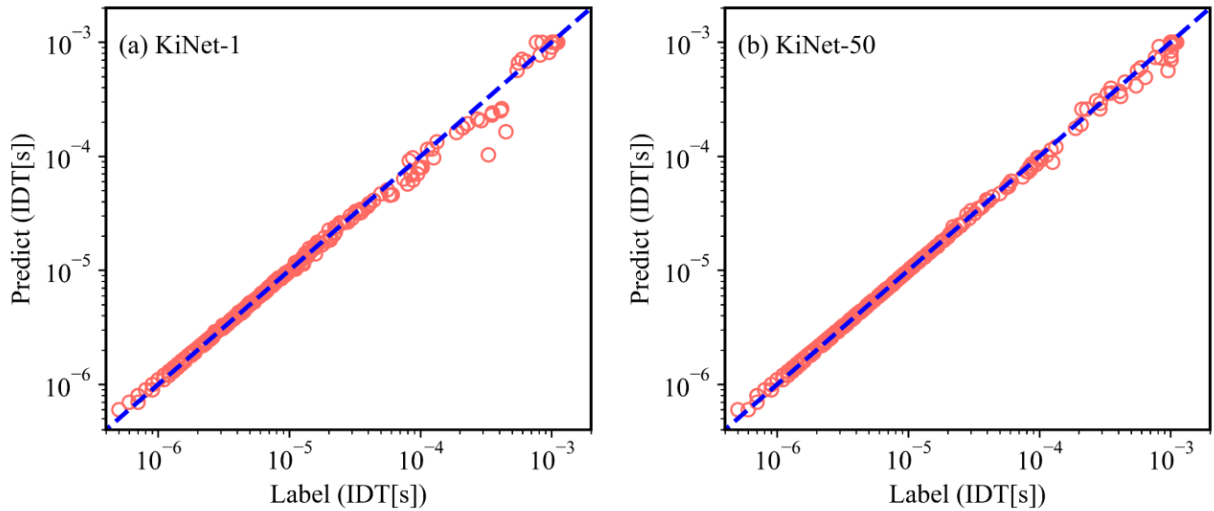

Figure 6. The predicted ignition delay time by (a) KiNet-1 and (b) KiNet-50 versus label data. 600 initial conditions are presented.

## 4. Efficient Computation of the Gradient of Ignition Delay Time

Next, we shall propose an efficient approach to compute the gradient of IDT to KiNet model parameters. As mentioned in the introduction, IDTs are easier to measure compared to the time-resolved species profiles and are widely available in the literature. Therefore, including the IDT as targets in addition to those available time-resolved species profiles during pyrolysis and ignition can greatly increase the amount of training datasets for KiNet. Computational studies with conventional kinetic models have shown that the gradient of IDT to the pre-factor A of the kinetic rate constant is parallel to the gradient of temperature at the ignition point [30,31]. In the present work, we shall investigate whether such correlation still holds for KiNet, since KiNet parameters not only depends on kinetic rates but also thermodynamic properties. We can efficiently compute the evolution of the gradient of temperature and species mole fractions to the KiNet model parameters with backpropagation. We denote the model parameters as a vector $\mathbf{W}$ and the gradient of a certain composition state as $\partial \Phi / \partial \mathbf{W}$. The magnitude of the gradient vector is given by $\|\partial \Phi / \partial \mathbf{W}\|_2$ and the direction is given by the unit vector $\frac{\partial \Phi / \partial \mathbf{W}}{\|\partial \Phi / \partial \mathbf{W}\|_2}$. The similarity between two directions can be measured by the cosine similarity between two gradient directions, i.e., the inner product between the corresponding unit vectors. The cosine similarity of unity and zero indicate that the two directions are perfect parallel and orthogonal, respectively.

Figure 7a shows the evolution of the gradient magnitude of temperature predicted by KiNet-50. It can be seen that the magnitude blows up near the ignition point, which is coinciding with the rapid temperature increasing. The behavior is also consistent with the kinetic sensitivity of rate constants shown in [30]. Figure 7b presents the evolution of the gradient directions for temperature and species mole fractions. All the cosine similarities are computed against the gradient direction of the temperature at the ignition point. Note that the gradient directions of temperature and species mole fractions converge to the same direction as approaching ignition. Such behavior was denoted as global similarity [32]. Previous study has shown that the global similarity is associated with the correlation between the gradient direction of species and that of IDT [30]. Therefore, we shall investigate if such correlation is still valid for KiNet.

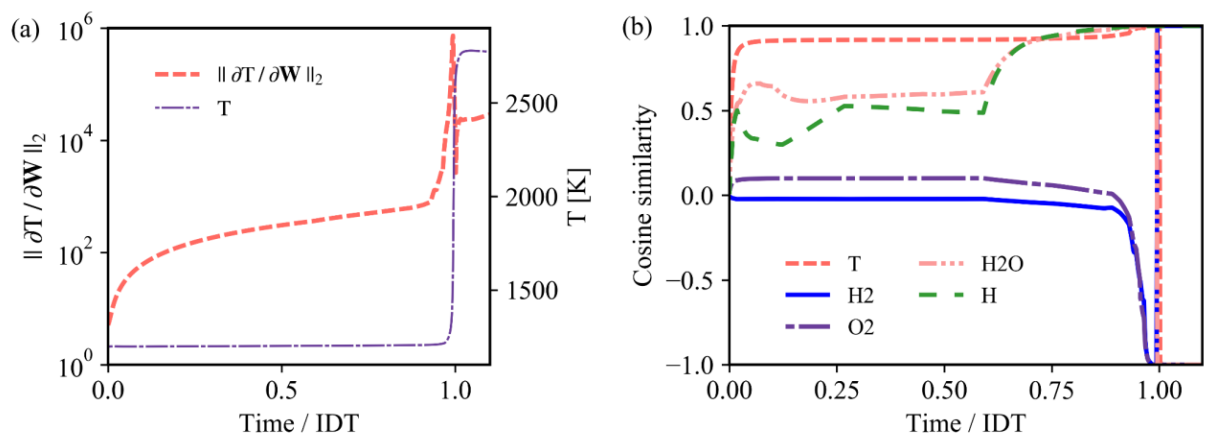

Figure 7. (a) The evolution of the magnitude and gradient magnitude of temperature. (b) The evolution of the cosine similarity evaluated between the gradient directions of temperature and species mole fractions and the gradient of the temperature at the instance of ignition. All the analysis is based on KiNet-50. The initial conditions are the same as in Fig. 2.

The gradient of IDT to the KiNet-50 model parameters is computed via the finite difference approach. The model contains a total number of 22,511 parameters. To alleviate the computational cost, only the gradient to the 11 parameters corresponding to the bias of the output layer is evaluated. Those 11 biases are the leading gradient components of the temperature at ignition point, contributing to 99.8% of the gradient magnitude, such that they are sufficient to represent the gradient direction of the IDT. The 11 components of the normalized gradients of IDT and temperature are then shown in Fig. 8, showing a very good consistency as the cosine similarity is 0.999. Therefore, the gradient direction of IDT for KiNet can be efficiently estimated with the gradient of temperature at the ignition point. The absolute value of the gradient of IDT can be readily acquired by computing the gradient of IDT to the most sensitive model parameter $w_m$ [30] via

$$\frac{\partial IDT}{\partial \boldsymbol{W}} = \frac{\partial T}{\partial \boldsymbol{W}} \frac{\partial IDT/\partial w_m}{\partial T/\partial w_m}. \tag{6}$$

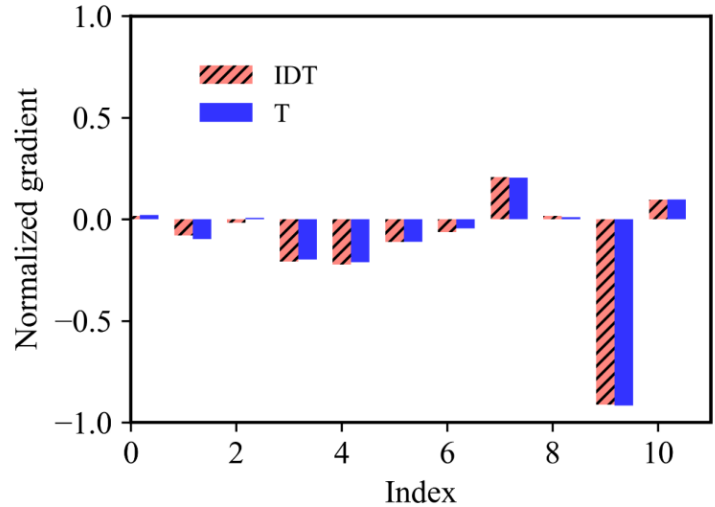

Figure 8. The normalized gradient of the ignition delay time and temperature at the ignition point to the 11 bias of the output layer. The cosine similarity between the two directions is 0.999. The initial conditions are the same as Fig. 2.

Finally, we demonstrate how to fine tune the trained KiNet with a new experimental measurement of IDT. The predicted IDT at the nominal condition in Fig. 2 by the originally trained KiNet-50 is 60.6 µs, and we shall update the KiNet with a measured IDT of 58.0 µs, for example. The loss function is given by

$$\text{Loss}(IDT) = \left(IDT^{predict} - IDT^{label}\right)^2. \tag{7}$$

The gradient of the loss with respect to the KiNet model parameter $\boldsymbol{W}$ would be

$$\frac{\partial Loss(IDT)}{\partial \mathbf{W}} = 2(IDT^{predict} - IDT^{label})\frac{\partial IDT}{\partial \mathbf{W}}, \qquad (8)$$

where the gradient of IDT is given by Eq. (6). The model is updated using gradient descent as

$$\mathbf{W}^{new} = \mathbf{W}^{old} - lr\frac{\partial Loss(IDT)}{\partial \mathbf{W}}, \qquad (9)$$

where the learning rate $lr$ is set to be $10^{-14}$. The training results after 100 epochs are shown in Fig. 9. The monotonically and smoothly decreasing loss of IDT shown in Fig. 9a confirms the effectiveness of the approximated gradient. The predicted temperature profiles using the old and updated model are then compared in Fig. 9b. The updated KiNet-50 not only accurately predicts the target IDT of 58.0 µs but also reasonably predicts the shape of the temperature profile.

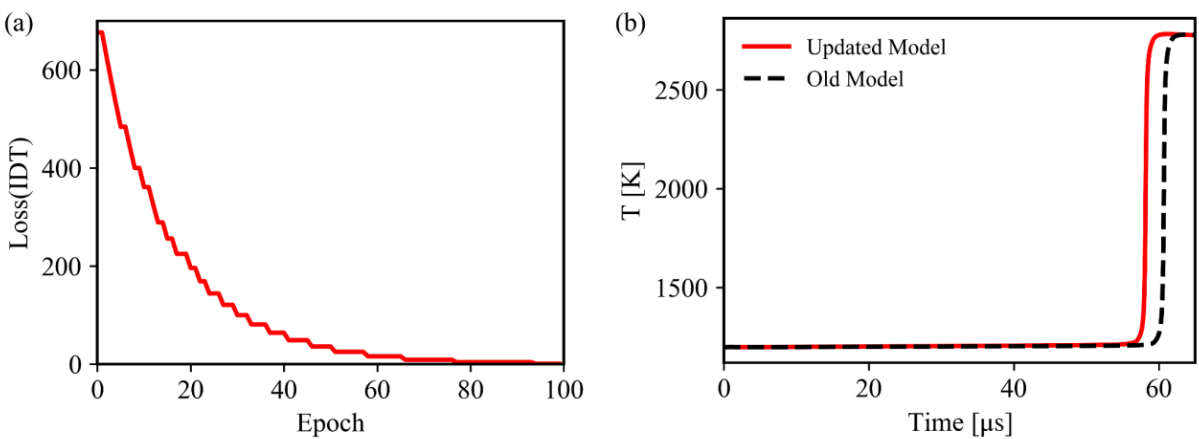

Figure 9. (a) The evolution of the loss of IDT during the update of KiNet-50 with an additional measurement of IDT and (b) the predicted temperature profiles using the old and updated KiNet-50. The initial conditions are the same as in Fig. 2.

## 5. Conclusions

We propose a DNN of KiNet for building deep learning representations of chemical kinetics. The KiNet incorporates recent advancement of (i) deep residual network (ResNet), (ii) backpropagation through time (BPTT), and (iii) the efficient computation of the gradient of the ignition delay time to kinetic model parameters. The training of KiNet is based on the simulated profiles of the composition states using a conventional $H_2$ kinetic model, and the trained KiNet can reproduce the time-resolved ignition process. Furthermore, we have demonstrated an efficient computation approach to backpropagate the loss of the predicted ignition delay time to the KiNet model parameters, which enables us to train KiNet leveraging the rich database of ignition delay time in the literature. The demonstration shall open up the possibility of building data-driven kinetic models autonomously. In addition, the trained KiNet could be potentially applied to kinetic model reduction and chemistry acceleration in turbulent combustion simulations.

## 6. Acknowledgment

SD would like to acknowledge the support from the d'Arbeloff Career Development allowance at Massachusetts Institute of Technology.